\documentclass[aps,prb,twocolumn,superscriptaddress]{revtex4}
\usepackage{graphicx}
\usepackage{amssymb}
\usepackage{amsmath}
\usepackage{graphicx}
\usepackage{epsfig}
\usepackage{color}
\usepackage{ulem}

\begin{document}
\title{Hybridization fluctuations in the half-filled periodic Anderson model}
\author{Danqing Hu}
\affiliation{Beijing National Laboratory for Condensed Matter Physics and Institute of Physics, Chinese Academy of Sciences, Beijing 100190, China}
\affiliation{University of Chinese Academy of Sciences, Beijing 100049, China}
\author{Jian-Jun Dong}
\affiliation{Beijing National Laboratory for Condensed Matter Physics and Institute of Physics, Chinese Academy of Sciences, Beijing 100190, China}
\affiliation{University of Chinese Academy of Sciences, Beijing 100049, China}
\author{Yi-feng Yang}
\email[]{yifeng@iphy.ac.cn}
\affiliation{Beijing National Laboratory for Condensed Matter Physics and Institute of Physics, Chinese Academy of Sciences, Beijing 100190, China}
\affiliation{University of Chinese Academy of Sciences, Beijing 100049, China}
\affiliation{Songshan Lake Materials Laboratory, Dongguan, Guangdong 523808, China}
\date{\today}

\begin{abstract}
Motivated by recent photoemission and pump-probe experiments, we report determinant Quantum Monte Carlo simulations of hybridization fluctuations in the half-filled periodic Anderson model. A tentative phase diagram is constructed based solely on hybridization fluctuation spectra and reveals a crossover regime between an unhybridized selective Mott state and a fully hybridized Kondo insulating state. This intermediate phase exhibits nonlocal hybridization fluctuations and consequentially the so-called ``band bending" and a direct hybridization gap as observed in angle-resolved photoemission spectroscopy and optical conductivity. This connects the band bending with the nonlocal hybridization fluctuations as proposed in latest ultrafast optical pump-probe experiment. The Kondo insulating state is only established at lower temperatures with the development of sufficiently strong inter-site hybridization correlations. Our work suggests a unified picture for interpreting recent photoemission, pump-probe, and optical observations and provides numerical evidences for the importance of hybridization fluctuations in heavy fermion physics.
\end{abstract}

\maketitle

Heavy fermion materials, mostly rare earth or actinide intermetallics, provide a model system for studying the localized-to-itinerant transition of strongly correlated electrons \cite{Stewart1984RMP,Hewson1997,Coleman2015,Onuki2018}. Theoretically, this transition is attributed to collective hybridizations between localized and conduction electrons \cite{Mott1974,Doniach1977,Coleman1983PRB,Yang2008nature}. A mean-field approximation has often been assumed with a static and  uniform hybridization \cite{Read1983JPC,Coleman1984PRB,Millis1987PRB,Newns1987AdvPhys}, leading to many interesting predictions \cite{Zhang2000PRB,Burdin2009PRB,Dzero2010PRL,Dubi2011PRL,Ramires2012PRL} and the identification of a characteristic coherence temperature separating the hybridized and unhybridized states \cite{Iglesias1997PRB,Burdin2000PRL,Assaad2002PRB}. The hybridization is manifested by a bending of the conduction bands that also marks the emergence of heavy electrons. However, this simple understanding was questioned by recent angled-resolved photoemission spectroscopy (ARPES) \cite{Chen2017PRB}, which revealed a ``band bending" well above the coherence temperature. Although transport and band properties may not have an exact microscopic correspondence, this separation between coherence and hybridization still caused some confusion on the conventional picture.

Fortunately, some light was shed on this issue lately by ultrafast optical pump-probe experiment \cite{Liu2019arXiv}, in which a two-stage hybridization scenario was proposed based on the analysis of anomalous quasiparticle relaxation. While the low-temperature stage starts at  the coherence temperature and results in a fluent-dependent relaxation associated with an indirect hybridization gap on the density of states as predicted by the mean-field theory, a precursor ungapped stage was also revealed to exhibit hybridization fluctuations whose onset temperature coincides with that of the ``band bending" in ARPES. Such a precursor stage is beyond the mean-field description and has not been sufficiently explored. Although it has been argued that hybridization fluctuations might play an important role in heavy fermion physics  \cite{Pepin2007PRL,Pepin2008PRB,Yang2017}, further studies have been largely hindered by difficulties in analytical treatment. This is unfortunate because hybridization fluctuations might be the basis of many important heavy fermion phenomena \cite{Andres1975PRL,Steglich1976PRL,Sigrist1991RMP,Stewart2001RMP,Gegenwart2008NatPhys,White2015}.

To avoid the analytical difficulties, we propose in this work to study hybridization fluctuations numerically using determinant Quantum Monte Carlo (DQMC)  \cite{Blankenbecler1981PRD,Assaad2008,Tomas2012IEEE}. DQMC has led to many useful insights on heavy fermion physics  \cite{Vekic1995PRL,Capponi2001PRB,Euverte2013PRB,Jiang2014PRB,Wei2017SciRep,Costa2019PRB,LufengZhang2019PRB}, but this issue has not been well discussed. Although the calculations are often limited at half filling to avoid the sign problem \cite{Loh1990PRB}, the exact numerical results will still allow us to extract some generic properties beyond the mean-field approximation. In particular, one may want to know if there are indeed multiple stages of hybridization as proposed in pump-probe experiment and how they might be connected with the ``band bending" in ARPES and the lattice coherence (here referring to the Kondo insulating state with a fully opened indirect hybridization gap as predicted in the mean-field theory). To this end, we constructed a tentative phase diagram based solely on hybridization fluctuation spectra. A partially hybridized precursor state was then revealed that exhibits low-energy hybridization fluctuations with the so-called ``band bending" in the dispersion, while the Kondo insulating state is only established at lower temperatures with sufficiently strong inter-site hybridization correlations. This confirms the two-stage hybridization scenario and suggested a consistent interpretation for the photoemission, pump-probe, and optical spectroscopies.

We start with the periodic Anderson model on a two-dimensional square lattice,
\begin{align}
H &  =-t\sum_{\langle ij\rangle\sigma}\left(c_{i\sigma}^{\dag
}c_{j\sigma}+c_{j\sigma}^{\dag}c_{i\sigma}\right)+V\sum_{i\sigma}(  f_{i\sigma}^{\dag}c_{i\sigma
}+h.c.)  \nonumber\\
&  \quad+E_{f}\sum_{i\sigma}f_{i\sigma}^{\dag}f_{i\sigma}+U\sum_{i}(n_{i\uparrow}^{f}-\frac{1}{2})  (  n_{i\downarrow}^{f}-\frac{1}{2}),
\end{align}
where $c_{i\sigma}^{\dag}(  c_{i\sigma})  $ and $f_{i\sigma}^{\dag}(  f_{i\sigma})  $\ are the creation (annihilation) operators of conduction and localized $f$ electrons, respectively. $t$ is the hopping integral of conduction electrons between nearest-neighbor sites, and $V$ is the bare hybridization. We set $t=1$ for the energy unit, $U=6$ for the Coulomb interaction of $f$ electrons, and $E_f=0$ for the particle-hole symmetry to avoid the sign problem in the Monte Carlo simulations.

To study hybridization fluctuations, we first introduce the hybridization field, $O_i=\sum_{\sigma}(c^\dagger_{i\sigma}f_{i\sigma}+f^\dagger_{i\sigma}c_{i\sigma})$, and define its correlation function,
\begin{equation}
L_{ij}(\tau)=-\left\langle \mathcal{T}_\tau \left[O_i(\tau)-\langle O_i\rangle\right]\left[O_j(0)-\langle O_j\rangle\right]\right\rangle,
\end{equation}
where $\mathcal{T}_\tau$ is the ordering operator for the imaginary time $\tau$. Unlike the Kondo lattice model, where a static hybridization is nothing but a mean-field artefact, the thermodynamic average $\langle O_i\rangle$ here is always finite and thus not a good quantity to distinguish the physically unhybridized and hybridized states \cite{Bernhard2000PRB}. It is therefore subtracted to highlight the dynamical hybridization fluctuations. The model is then evaluated numerically with DQMC \cite{Blankenbecler1981PRD,Assaad2008,Tomas2012IEEE}. The imaginary time is discretized into $M$ slices with the inverse temperature $\beta=M\Delta \tau$. At each site and time slice, the interaction is decoupled using the Hubbard-Stratonovich transformation by introducing an auxiliary Ising field. The resulting bilinear Hamiltonian can be treated exactly and the correlation function can be calculated with the help of Wick's theorem before averaging over all sampled field configurations. Our simulations were performed on an 8$\times$8 square lattice with $M=80$ and examined with larger lattice size and time slices. The hybridization spectral function, $A_{\mathbf{q}}( \omega)  =-\frac{1}{\pi}\operatorname{Im}L_{\mathbf{q}}( \omega)$, was solved using the maximum entropy method for
\begin{equation}
L_{\mathbf{q}}(  \tau)  =\int_{-\infty}^{\infty}{\text d}\omega\frac {e^{-\tau\omega}}{e^{-\beta\omega}-1}A_{\mathbf{q}}(  \omega),
\end{equation}
where $L_{\mathbf{q}}(  \tau)  =\frac{1}{N}\sum_{ij}e^{-i\mathbf{q}\cdot( \mathbf{r}_{i}-\mathbf{r}_{j})  }L_{ij}(  \tau)$. The real part of  $L_{\mathbf{q}}(  \omega)  $ was then calculated using the Kramers-Kronig relation. To the best of our knowledge, these quantities have not been well explored in previous studies. The fermionic spectral functions were also calculated following similar standard procedures for comparison.

\begin{figure}[ptb]
\begin{center}
\includegraphics[width=0.48\textwidth]{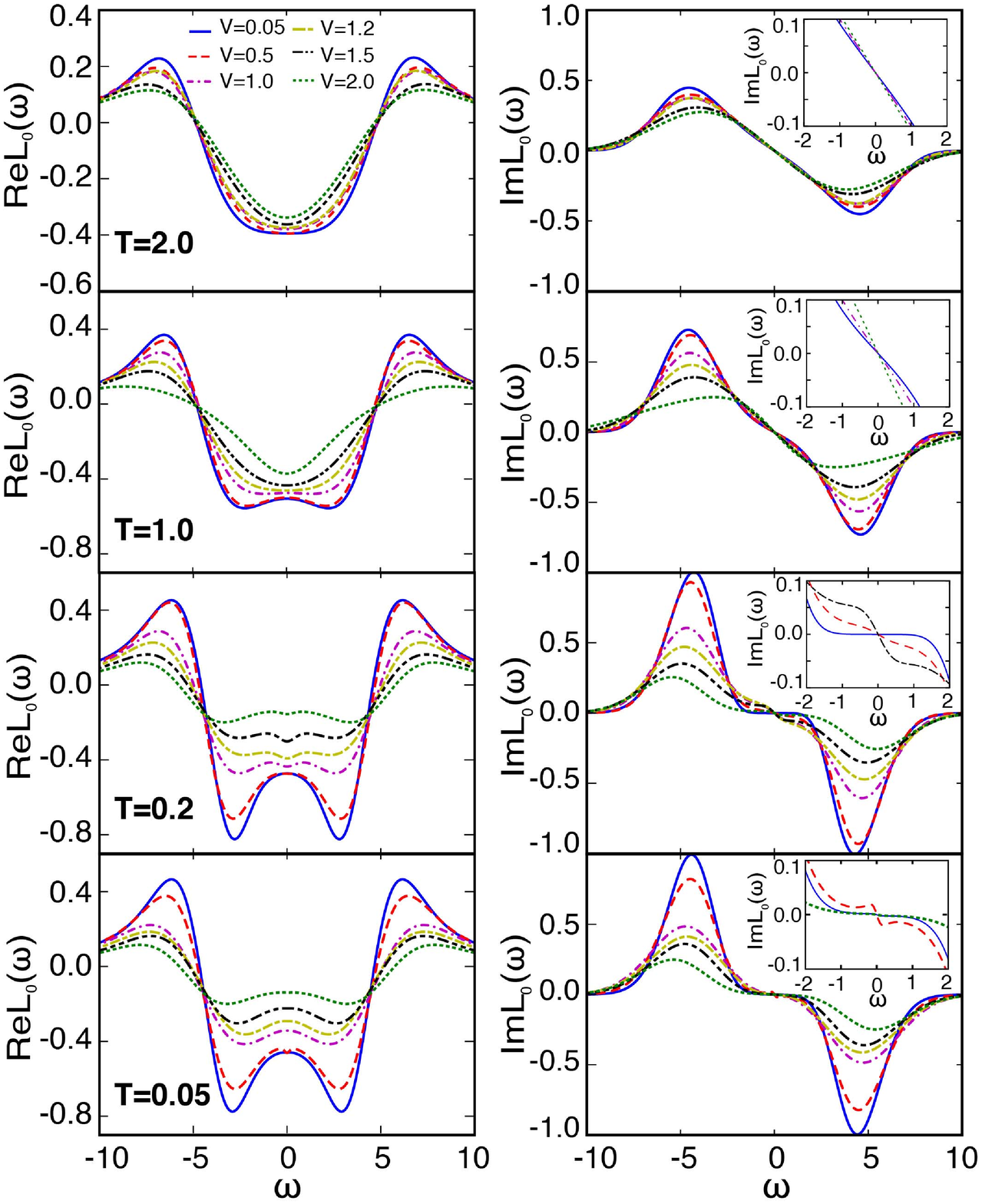}
\end{center}
\caption{(Color online) The real and imaginary parts of the hybridization correlation function, $L_{\mathbf{0}}(\omega)$, with the hybridization parameter $V$ and temperature $T$. The insets are enlarged plots of the imaginary part around $\omega=0$, showing the variation of the low-energy slope in $\operatorname{Im} L_{\mathbf{0}}(\omega)$ with different parameters.}
\label{fig1}
\end{figure}

Figure~\ref{fig1} plots the real and imaginary parts of $L_{\mathbf{q=0}}(  \omega)$ for varying $V$ at different temperatures. We first consider the high temperature regime. For $T=2.0$, $L_{\mathbf{0}}(\omega)$ changes only slightly with $V$ and shows two peaks at $\omega\approx\pm U$ due to excitations between two $f$ electron Hubbard bands. The finite slope in $\operatorname{Im} L_{\mathbf{0}}(\omega)$ around $\omega=0$ persists for $V=0$ (not shown) and must result from thermal excitations of unhybridized $f$ and conduction electrons. For $T=1.0$ and small $V$, the single valley in $\operatorname{Re} L_{\mathbf{0}}(\omega)$ evolves into a small hump with two valleys at $\omega\approx \pm U/2$, indicating the suppression of thermal excitations with lowering temperature. The two-valley features can be understood from the $V=0$ limit, where the correlation function has an analytical form,
\begin{equation}
L_{\mathbf{0}}(  \omega)=\frac{2}{N}\sum_{\mathbf{k},\alpha=\pm}\frac{[  f(\alpha U/2)  -f( \epsilon_\mathbf{k}) ]  (\epsilon_\mathbf{k}-\alpha U/2)  }{(  \omega+\operatorname*{i}\eta)  ^{2}-( \epsilon_\mathbf{k}-\alpha U/2)  ^{2}},
\label{eqV0}
\end{equation}
where $f(x)$ is the Fermi distribution function and $\eta=0^+$ is an infinitesimal cutoff. For a flat band with a half bandwidth $D$, the summation over $\mathbf{k}$ can be evaluated exactly at zero temperature and yield, $L_{\mathbf{0}}\left(  \omega\right)  =D^{-1}\ln\frac{\left( \omega+\operatorname*{i}\eta\right)  ^{2}-\left( U/2\right)  ^{2}}{\left( \omega+\operatorname*{i}\eta\right)  ^{2}-\left(  D+U/2\right)  ^{2} }$, which explains the calculated minima and maxima around $\omega=\pm U/2$ and $\pm (D+U/2)$. For large $V$, however, the single-valley shape is recovered.

The two-valley feature can be seen more clearly at $T=0.2$. Correspondingly, the low-energy slope in $\operatorname{Im} L_{\mathbf{0}}(\omega)$ becomes almost zero, indicating diminishing thermal excitations. However, for larger $V$, a small dip appears around $\omega=0$ on top of the hump in $\operatorname{Re} L_{\mathbf{0}}(\omega)$. Accordingly, the imaginary part exhibits a large slope in a small low-energy window followed by a sharp kink before turning to a high-energy plateau above $|\omega|\approx 0.2$. The finite slope must be a quantum effect and indicates a regime with low-energy hybridization fluctuations due to the coupling between conduction and $f$ electrons. Similar features can be found at $T=0.05$ for $V=0.5$, but are suppressed at larger $V$, where the dip in $\operatorname{Re} L_{\mathbf{0}}(\omega)$ is filled in and turns into a smooth maximum, and the slope in $\operatorname{Im} L_{\mathbf{0}}(\omega)$ is also suppressed.

\begin{figure}[t]
\begin{center}
\includegraphics[width=0.48\textwidth]{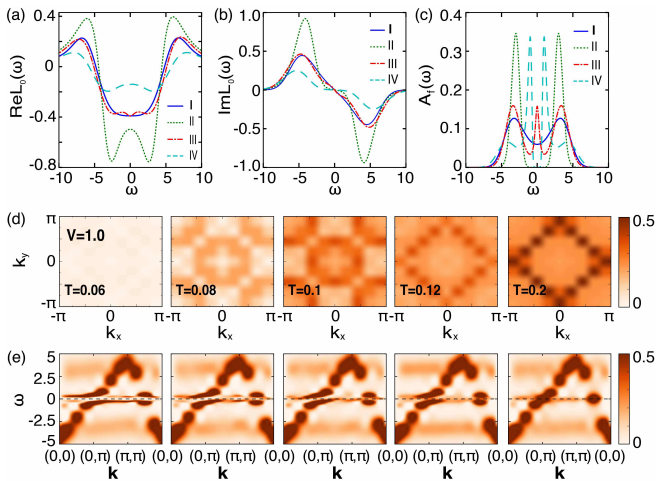}
\end{center}
\caption{(Color online) (a, b) Comparison of the different features of the real and imaginary parts of $L_{\mathbf{0}}(  \omega)$ in four distinct regimes; (c) The corresponding $f$ electron local density of states. The parameters are: $T=2.0$, $V=0.5$ for regime I; $T= 0.6$, $V=0.25$ for regime II; $T= 0.2$, $V=1.2$ for regime III; and $T = 0.05$, $V=2.0$ for regime IV. (d) Intensity plots of the fermionic spectral function at the Fermi energy in the Brillouin zone evolving with temperature for a fixed $V = 1.0$; (e) The corresponding plots of the dispersion along a chosen path in the Brillouin zone, showing its evolution across the intermediate regime III.}
\label{fig2}
\end{figure}

The above distinct features of $L_{\mathbf{0}}(\omega)$ suggest four different regimes of the periodic Anderson model. The results are summarized in Figs.~\ref{fig2}(a) and \ref{fig2}(b). Regimes I and II are governed by background contributions of decoupled conduction and $f$ electrons. For regime I, thermal excitations are large such that the real part of $L_{\mathbf{0}}(\omega)$ has only one valley and the imaginary part has a finite slope; while for regime II, thermal effects are suppressed, revealing two valleys at $\omega=\pm U/2$ in $\operatorname{Re} L_{\mathbf{0}}(\omega)$ due to the Hubbard bands, and the slope in $\operatorname{Im} L_{\mathbf{0}}(\omega)$ is consequentially reduced. The latter corresponds to a selective Mott regime of $f$ electrons that are effectively decoupled from conduction electrons. To see this, we plot the $f$ electron local density of states (DOS) in Fig.~\ref{fig2}(c). The spectra are governed by two broad Hubbard peaks at $\omega=\pm U/2$. In regime I, the valley in between is partially filled by thermal excitations, but in regime II, it is depleted and reveals the Mott gap \cite{Held2000PRL,Logan2016JPCM}.

\begin{figure}[t]
\begin{center}
\includegraphics[width=0.48\textwidth]{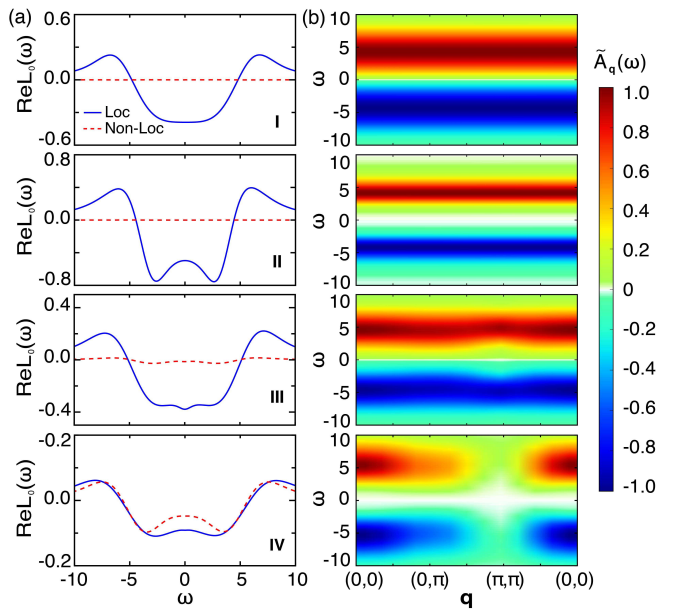}
\end{center}
\caption{(Color online) (a) Comparison of the local and nonlocal contributions to $\operatorname{Re} L_{\mathbf{0}}(\omega)$, showing the onset of nonlocal term in regime III and development in regime IV. (b) The normalized hybridization spectral function, $\tilde{A}_{\mathbf{q}}( \omega)$, showing the growth of hybridization correlations at $(\pi,\pi)$. The parameters are the same as in Fig.~\ref{fig2}(a).}
\label{fig3}
\end{figure}

Deviation from the above Mott features defines two hybridized regimes. The small dip on the hump of $\operatorname{Re} L_{\mathbf{0}}(\omega)$ and the large low-energy slope of $\operatorname{Im} L_{\mathbf{0}}(\omega)$ in regime III mark a genuine quantum effect due to low-energy  hybridization fluctuations. In regime IV, these features are again suppressed, indicating the crossover into a different phase. This is the Kondo insulating regime, where the slave-boson mean-field theory predicts an artificial boson condensation. The hybridization correlation function can also be evaluated analytically,
\begin{equation}
L_{\mathbf{0}}(  \omega)  =\frac{4}{N}\sum_{\mathbf{k}}\frac{f(  E_{\mathbf{k}-})  -f(  E_{\mathbf{k}+}) }{(\omega+\operatorname*{i}\eta)^2-\Delta_{\mathbf{k}} ^{2}}\frac{\epsilon_{\mathbf{k}}^2}{\Delta_{\mathbf{k}}},
\label{eqU0}
\end{equation}
where $E_{\mathbf{k}\pm}=(\epsilon_{\mathbf{k}}  \pm \Delta_{\mathbf{k}})/2$ denote two hybridization bands and $\Delta_{\mathbf{k}}=\sqrt{ \epsilon_{\mathbf{k}} ^{2}+\Delta_0^{2}}$ is the direct hybridization gap at each $\mathbf{k}$ with an effective hybridization strength $\Delta_0$ whose magnitude separates the hybridized and unhybridized phases. Thus $\operatorname{Re} L_{\mathbf{0}}(\omega)$ has two minima around $\omega=\pm\Delta_0$. Since $\Delta_{\mathbf k}\ge\Delta_0$ for all $\mathbf{k}$, we have the imaginary part, $\operatorname{Im} L_{\mathbf{0}}(\omega)\propto\sum_{{\mathbf k},\alpha=\pm}  \alpha\delta(\omega+\alpha\Delta_{\mathbf k})\epsilon_{\mathbf k}^2/\Delta_{\mathbf k}^2 $, which is gapped for $|\omega|<\Delta_0$. Obviously, the above formula fails in regime III, where we have a large low-energy slope in $\operatorname{Im} L_{\mathbf{0}}(\omega)$ due to the presence of hybridization fluctuations. This has an immediate consequence on the $f$ electron spectra. As shown in Fig.~\ref{fig2}(c), instead of a Kondo insulating gap in the local DOS as in regime IV, we find a broad peak around $\omega=0$, making regime III a precursor ungapped state beyond the mean-field approximation; while in regime IV, the two sharp peaks at lower energyies can be roughly understood from the band hybridization in $E_{\mathbf{k}\pm}$.

To gain further insight, we plot in Fig.~\ref{fig2}(d) the momentum distribution of the total fermionic ($f$ and conduction electrons) spectral intensity at the Fermi energy evolving with temperature for $V=1.0$. We see a clear crossover from a selective Mott regime with two Hubbard bands and a small conduction electron Fermi surface to a Kondo insulating regime where the hybridization gap is fully opened with no discernible spectral weight at the Fermi energy in the whole Brillouin zone. In between, regime III shows a finite spectral weight (not the Fermi surface), albeit with a very different pattern. For clarity, we plot the dispersion in Fig.~\ref{fig2}(e), where a slight band bending is already seen in regime III, but the gap is only partially opened, leaving a finite spectral weight at the Fermi energy and the broad peak in the local DOS. This agrees with the ARPES observation \cite{Chen2017PRB} and supports the two-stage scenario proposed by pump-probe experiment \cite{Liu2019arXiv}. The band bending is also an indication of the direct hybridization gap as probed in optical conductivity \cite{Chen2016RPP}. This gives a consistent interpretation of the high-temperature features in ARPES, pump-probe and optical measurements.

To understand how hybridization fluctuations can further induce the $f$ electron coherence (here the Kondo insulating state) at lower temperature, we compare in Fig.~\ref{fig3}(a) the local and nonlocal contributions to $\operatorname{Re}L_{\mathbf{0}}(\omega)$. Since $L_{\mathbf{0}}(\omega)=N^{-1}\sum_{ij}L_{ij}(\omega)$, the nonlocal part is a sum of all inter-site correlations. We see for regimes I and II, the nonlocal contribution is indiscernible. It only starts in regime III but, quite surprisingly, becomes comparable with the local one in regime IV. Its very  existence is an indication of quantum effect. Clearly, while the band bending already appears in regime III, the lattice coherence can only be established later with sufficiently strong inter-site hybridization correlations. It should be noted that the nonlocal correlation is  dominantly contributed by the nearest-neighbor term in our calculations. Hence, the Kondo insulator should be viewed more like a short-range-correlated insulator rather than a simple band insulator described by the mean-field picture. This short-range correlation is consistent with previous calculations as well as nuclear magnetic resonance (NMR) observations on doped Kondo lattice \cite{Wei2017SciRep,Lawson2019}. We further remark that the development of nonlocal correlations is also manifested in the momentum space. Figure \ref{fig3}(b) plots the normalized hybridization spectral function, $\tilde{A}_{\mathbf{q}}( \omega)$, along the path $(0,0)-(0,\pi)-(\pi,\pi)-(0,0)$ in the Brillouin zone. The spectra are basically featureless besides the Hubbard bands in regimes I and II. A slight change appears at $(\pi,\pi)$ in regime III, which grows rapidly in regime IV and eventually intrudes into the Mott feature. This reflects a competition between nonlocal hybridization correlations and the local Mott physics. The fact that the former emerges dominantly near $(\pi,\pi)$ seems to also indicate an interplay between hybridization and magnetic fluctuations \cite{Yang2017}.

\begin{figure}[t]
\begin{center}
\includegraphics[width=0.47\textwidth]{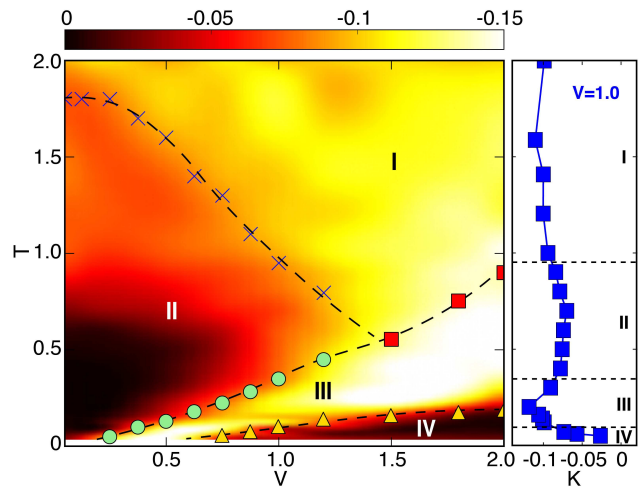}
\end{center}
\caption{(Color online) A tentative phase diagram constructed based solely on $L_{\mathbf{0}}(  \omega)$. The points are estimated from the different features of its real part and the lines are a guide to the eye. The background colors reflect the low-energy slope $K$ of its imaginary part. The right panel plots the values of $K$ for $V = 1.0$, whose nonmonotonic temperature dependence clearly demonstrates the separation of four regimes.}
\label{fig4}
\end{figure}

Putting together, we find it possible to construct a tentative phase diagram of the periodic Anderson model based solely on hybridization fluctuation spectra. The result is shown in Fig.~\ref{fig4}, where the points and dashed lines mark the phase (crossover) boundaries extracted roughly from the features of $\operatorname{Re} L_{\mathbf{0}}(\omega)$, and the background colors reflect the magnitude of the slope, $K=\left.\mathrm{d}\operatorname{Im} L_{\mathbf{0}}(  \omega) /\mathrm{d}\omega\right|_{\omega=0}$. We see a rough agreement between the two methods. The phase diagram reveals clearly the four distinct regimes and their overall relationship. This is best demonstrated in the right panel of Fig.~\ref{fig4} for $V=1.0$, where $K$ undergoes a nonmonotonic variation that separates the different regimes. It is now evident that regime III (at small $V$) bridges the unhybridized selective Mott state (II) and the fully hybridized Kondo insulating state (IV) and marks a crossover from localized to itinerant $f$ electrons. In previous analytical calculations, it has been proposed that the localized-to-itinerant transition at zero temperature may be viewed as a selective Mott transition \cite{Pepin2007PRLmott,Pepin2008PRBmott}. This seems to be consistent with our results if regime III could  in some way be associated with the crossover regime above the Mott critical end point. Unfortunately, at the moment our calculations are limited at relatively higher temperatures and it is not clear if a straightforward connection can be made. We should note that the presence of a precursor regime above the Kondo insulating phase can also be seen in previous calculations \cite{Jarrell1993PRL,Medici2005PRL}, but it has not been well discussed in the context of hybridization fluctuations. It will be important if our study can be extended to extremely low temperatures to provide numerical evidences for previous analytical treatment. Recently, it has also been proposed that non-Hermitian physics might lead to exotic properties in a Kondo insulator \cite{Shen2018PRL,Yoshida2018PRB,Michishita2019arXiv}. The so-called exceptional points were argued to be around the high-temperature boundary of the Kondo insulating phase \cite{Michishita2019arXiv}. In our case, if we make the replacement $\eta\rightarrow \Gamma_k$ in Eq.~(\ref{eqU0}), we will be able to get a finite slope, $K\propto -\sum_{\mathbf{k}}  \Gamma_\mathbf{k}\epsilon_{\mathbf{k}}^2/\Delta_{\mathbf{k}}(\Gamma_\mathbf{k}^2+\Delta_\mathbf{k}^2)^2$, which approaches zero when $\Gamma_{\bf k}\rightarrow 0$ or $\infty$. Thus the finite $K$ in regime III might be associated with the finite dissipation (or lifetime) of hybridization or fermionic excitations in the crossover phase. It would certainly be more intriguing if regime III is a state that could potentially host some exotic non-Hermitian physics.

To summarize, we studied hybridization fluctuations with DQMC for the half-filled periodic Anderson model. This allows us to extract some useful information beyond the mean-field approximation and construct a tentative phase diagram based solely on hybridization fluctuation spectra. We found a crossover from an unhybridized selective Mott state to a fully hybridized Kondo insulating state. In between, there exists an intermediate phase with low-energy hybridization fluctuations and evident band bending. The $f$ electron coherence is only established at lower temperatures with the development of sufficiently strong inter-site hybridization correlations. This confirms the proposed two-stage hybridization scenario based on recent ARPES and pump-probe experiments. The band bending occurs first near the Fermi wave vector of conduction electrons and gives rise to a direct hybridization gap as probed in optical conductivity well above the coherence temperature. We have thus a consistent picture for the high-temperature features of photoemission, pump-probe, and optical spectroscopies. Possible connections with Mott and non-Hermitian physics were  also discussed briefly. Our work provides a promising start for numerical studies of hybridization dynamics in causing exotic correlated properties of heavy fermion systems. In the future, we expect to see more insights if our study could be extended to the quantum critical regime or the metallic phase away from the half filling to make a full comparison with previous analytical or experimental conclusions.

This work was supported by the National Natural Science Foundation of China (NSFC Grant No. 11974397,  No. 11522435), the National Key R\&D Program of China (Grant No. 2017YFA0303103), the State Key Development Program for Basic Research of China (Grant No. 2015CB921303), the National Youth Top-notch Talent Support Program of China, and the Youth Innovation Promotion Association of CAS.

\end{document}